\documentclass{pasj00}
\begin{document}

\SetRunningHead{K. Torii et al.}{X-ray Spectrum of G330.2+1.0}
\Received{2005/10/30}
\Accepted{2005/12/16}

\title{Discovery of a Featureless X-Ray Spectrum in the Supernova Remnant Shell of G330.2+1.0}

\author{Ken'ichi \textsc{Torii}, Hiroyuki \textsc{Uchida}, Kazuto \textsc{Hasuike}, and Hiroshi \textsc{Tsunemi}%
}
\affil{Department of Earth and Space Science, Graduate School of Science, Osaka
 University\\ 1-1 Machikaneyama-cho, Toyonaka, Osaka 560-0043}
\author{Yasuhiro \textsc{Yamaguchi} and Shinpei \textsc{Shibata}}
\affil{Department of Physics, Yamagata University,\\ Kojirakawa, Yamagata, Yamagata 990-8560}


\KeyWords{ISM: supernova remnants ---  acceleration of particles --- X-rays: individual (G330.2+1.0) } 

\maketitle

\begin{abstract}

We report here on the first pointed X-ray observation of the supernova
remnant (SNR) G330.2+1.0. The X-ray morphology is characterized by an extended
shell. Its X-ray spectrum is well represented by a single power-law
function with a photon index of $\gamma\simeq 2.8$ and interstellar
absorption of $n_{\rm H}\simeq2.6\times 10^{22}$[cm$^{-2}$]. We interpret
this emission as synchrotron radiation from accelerated electrons at
the SNR shock, as seen in SN 1006. The surface brightness of the X-ray
emission is anti-correlated with the radio emission, and the power-law
spectrum is dominated at the western shell where the radio emission is
weak.  The co-existence of two distinct (radio bright/X-ray faint and
radio faint/X-ray bright) shells in a single supernova remnant
challenges our understanding of the particle acceleration and
radiation mechanisms in different interstellar environments. The
object may be a good target for searching TeV gamma-rays and molecular
gas surrounding the blast shock. We also report on the nature of a
bright point-like source (AX~J1601$-$5143) to the south of the SNR.

\end{abstract}

\section{Introduction}

 Our understanding of the origin and the acceleration mechanism of
cosmic-ray particles has been a great challenge in astrophysics
since the last century. X-ray observations have revealed that
acceleration is taking place in supernova remnant shells by
detecting featureless power-law type spectra (Koyama et al. 1995;
Koyama et al. 1997; Slane et al. 2001). These spectra are
interpreted as synchrotron radiation from accelerated electrons. The
detections of TeV photons in RX~J1713.7$-$3946 (Muraishi et al. 2000; Aharonian
et al. 2004) and RX J0852.0$-$4622 (Aharonian et al. 2005) have assured the
presence of high-energy ($>$TeV) particles, either leptons or hadrons,
in these SNR shells.

 The radio source of G330.2+1.0 is characterized by a clumpy distorted shell of
\timeform{11'} diameter (Caswell et al. 1983; Whiteoak, Green 1996). Caswell
et al. (1983) attributed its morphology as being due to the interaction of a
shock wave with a dense medium, particularly toward the lower galactic
latitude (southeast). A clumpiness in the radio map led Whiteoak and
Green (1996) to classify G330.2+1.0 as a possible composite type. 
As we show in the following sections, there is no signature of
a Crab-like, compact X-ray source at the position of the clumpy radio
shell. We therefore classify G330.2+1.0 as a shell type. 

 The H\emissiontype{I} absorption spectrum to the SNR gives a minimum kinematic distance
of $4.9\pm 0.3$ kpc, while the greater distance for velocities interior to
the solar circle, 9.9~kpc, can not be excluded (McClure-Griffiths et
al. 2001; figure 4 and table 2). A possible association of the SNR with
the surrounding H\emissiontype{I} shell (McClure-Griffiths et al. 2001; subsection
5.3.2 and figure 14) favors that the SNR is close to the absorbing H\emissiontype{I}
clouds at $4.9\pm0.3$~kpc (or at $9.9$~kpc). It is worth noting that the line of sight of
$l=$\timeform{330D.2} intersects the Norma arm at distances of $\simeq6.3$~kpc
and $\simeq11.3$~kpc (McClure-Griffiths et al. 2001; figure 14). Therefore, a reasonable distance upper limit to the SNR may be
$11.3$~kpc. Hereafter, we denote the SNR distance as $d_{4.9}$
normalized to 4.9~kpc.  For this distance and the radio shell's
radius, the Sedov solution's age--radius relation gives a minimum age
of $t=3.1\times 10^{3}\, (E_{51}/n_{0})^{-1/2}\, d_{4.9}^{5/2}$ yr, with an explosion energy normalized to
$10^{51}$erg and the interstellar medium density normalized to 1\, ${\rm cm^{-3}}$.

\section{Observations}

 Observations of SNR G330.2+1.0 were performed with the ASCA
 (Tanaka et al. 1994) on 1999 September 11--12. ASCA had four X-ray telescopes that were equipped with two CCD cameras (SIS~0 and SIS~1) and two imaging gas-scintillation proportional counters (GIS~2 and GIS~3) (Ohashi et al. 1996; Makishima et al. 1996).
The SIS was operated in the 1 CCD mode and its field of view
 (\timeform{11'}$\times$\timeform{11'}) did not cover the full spatial extent of the SNR
 shell. We therefore concentrated on the GIS data, which has a larger
 field of view (\timeform{50'} diameter) and a higher detection efficiency for  hard X-ray
 sources. The effective exposure time for each GIS was 19.6~ks.

\section{Analysis and Results}

\begin{figure}
  \begin{center}
    \FigureFile(80mm,80mm){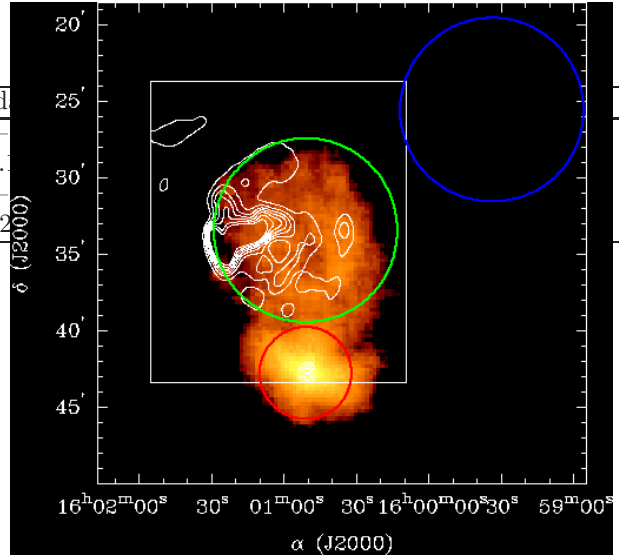}
  \end{center}
  \caption{X-ray image of the SNR shown by half tone. The contours show the radio map observed with the MOST (Whiteoak, Green 1996). 
The southern, middle, and northern circles show the regions from which
the X-ray spectra of AX~J1601$-$5143, G330.2+1.0, and the background,
respectively, were extracted.  }\label{fig1}
\end{figure}

\begin{figure}
  \begin{center}
    \FigureFile(70mm,70mm){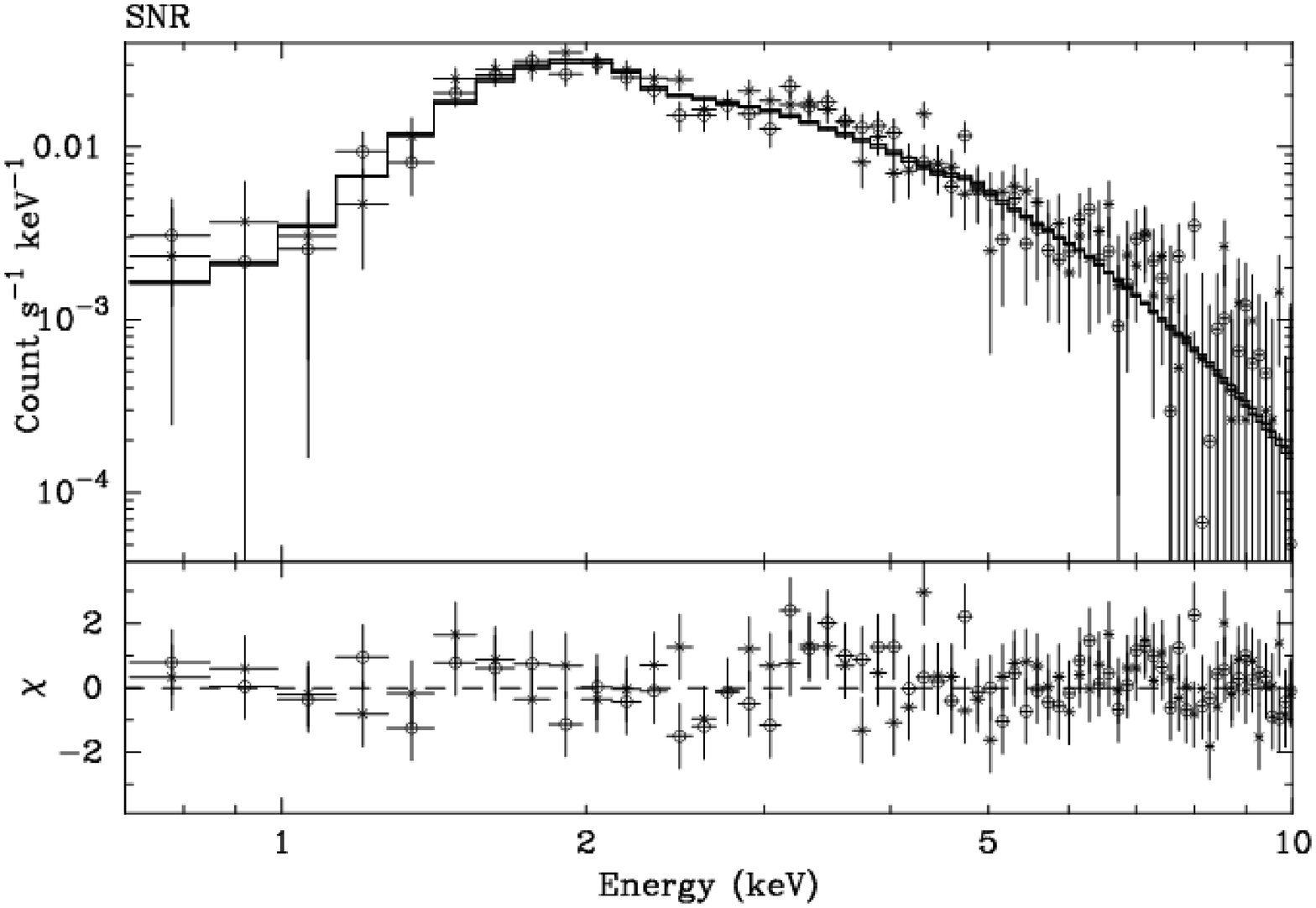}
    \FigureFile(70mm,70mm){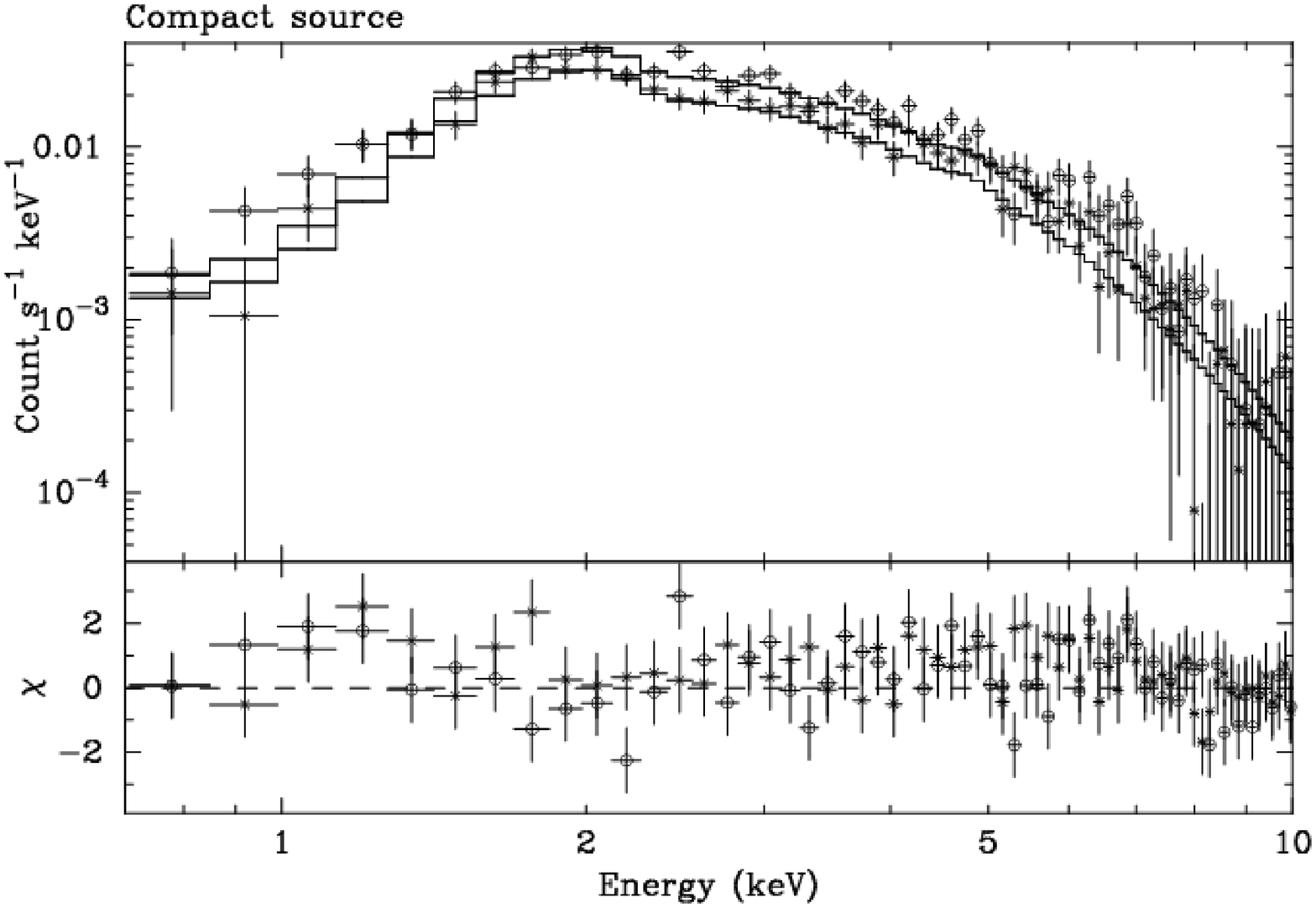}
  \end{center}
  \caption{X-ray spectra for the SNR and the compact source shown with the best-fit power-law function modified by the absorption. The circles and crosses show GIS~2 and GIS~3 data points, respectively.
}\label{fig2}
\end{figure}

Figure~\ref{fig1} shows an X-ray image of G330.2+1.0 obtained with the GIS.
An extended shell is clearly detected as well as a compact source in the
south.  The position of the compact source, as determined from the GIS
image, is (16:00:51.0, $-$51:42:33)(J2000) (hereafter AX~J1601$-$5143) with a
possible error of a few arcminutes. The radial profile of the compact
source is found to be consistent with a point source.  Excluding
background, the GIS count rate for AX~J1601$-$5143 was 0.094 cps and 0.074 cps
for GIS~2 and GIS~3, respectively. The lightcurve of AX~J1601$-$5143 does not show any significant flux variability from timescales of minutes to hours. We did
not detect any coherent pulsations either up to the Nyquist frequency of
128~Hz.
Figure~\ref{fig2} shows the GIS image overlaid on the radio map. The
radio emission is brightest at the eastern shell (toward the lower
galactic latitude), while the X-ray emission is brighter in the western part.
The surface brightness of the shell is clearly anti-correlated between
the radio and X-ray bands. 

Figures~\ref{fig2}a and \ref{fig2}b show the GIS spectra for the SNR
shell and the compact source, respectively, fitted with a
power-law function modified by the interstellar absorption. As shown
in figure~\ref{fig1}, circular regions of $6'$ and $3'$ radii, centered at
(16:00:50.9, $-$51:33:27)(J2000) and (16:00:50.9, $-$51:42:47)(J2000),
were used to extract spectra for the SNR and the compact source,
respectively. Background spectra were extracted from a circular region
of \timeform{7'.5} radius centered at (15:59:33.7, $-$51:25:35)(J2000).
Spectral fits to a non-thermal (power-law) model and thermal ({\tt mekal})
model by using XSPEC (Arnaud, 1996) are summarized in table \ref{tab:spectral_parameter}.  We
further tried to investigate the spatial variation of the X-ray
spectra across the SNR shell, but limited statistics did not allow us
to find any significant variation.

\begin{table}
  \caption{Spectral parameters.}\label{tab:spectral_parameter}
  \begin{center}
    \begin{tabular}{ccccccc} \hline
Object & Model& Photon Index/kT & Abundance &$n_{\rm H} (10^{22}\, {\rm cm^{-2}})$& $\chi^2$/dof &Flux$^*$ \\  \hline
G330.2+1.0 & power-law & 2.82 (2.61--3.04) &---&2.58 (2.24--2.94) &327.5/391& $1.6\times 10^{-11}$\\
G330.2+1.0 & mekal &  2.99 (2.53--3.50) & $<0.17$ & 1.85 (1.62--2.12)&327.6/390& $9.1\times 10^{-12}$\\
AX~J1601$-$5143 & power-law &2.63 (2.47--2.82) & --- & 2.80 (2.51--3.11)&404.9/391& $3.0\times 10^{-11}$\\
AX~J1601$-$5143 & mekal & 3.32 (2.91--3.85) & 0.14(0.02--0.28)& 2.16 (1.94--2.38)&396.0/390& $1.9\times 10^{-11}$\\ \hline
\end{tabular}
$^*$ Unabsorbed $0.7-10$~keV flux in ${\rm erg\, s^{-1}\, cm^{-2}}$.
  \end{center}
\end{table}

\section{Discussion}

Since the absorption columns to G330.2+1.0 and AX~J1601$-$5143 (table
\ref{tab:spectral_parameter}) do not exclude a possibility that they
are at a similar distance, we explore if the two objects are
physically associated or not.  AX~J1601$-$5143 is probably the same object as
1RXS~J160045.0$-$514307 in the ROSAT all-sky survey faint source catalog
(Voges et al. 2000).  The cataloged position is $65''$ from the GIS
position. The PSPC countrate, $0.032\pm0.012$~cps, is as expected from
the GIS spectral parameters with either the mekal model or
the power-law model, taking into account the statistical uncertainties
of the PSPC and GIS data.

If the power-law model is appropriate, AX~J1601$-$5143 could be a Crab-like
object or a background active galactic nucleus. The absence of any 
significant long-term flux variability favors the former
interpretation, but the required transverse velocity, $v_{\rm t} = 3.9\times
10^{3}(t_{\rm age}[{\rm yr}]/3100)^{-1}\,d_{4.9}\, {\rm km\, s^{-1}}$, for the projected
angular distance from the shell center, $\simeq 8'.8$ or 13\,
$d_{4.9}$~pc, is unreasonably large.  Also, the observed photon index,
2.63, is steeper than any object powered by an energetic pulsar (e.g.,
Gotthelf 2003).  Thus, it is unlikely that AX~J1601$-$5143 is a Crab-like
pulsar/nebula.

If the thermal interpretation is correct, AX~J1601$-$5143 could be a nearby
($d<4.9$~kpc) white dwarf binary with $L_{\rm X}<5.5\times
10^{34}\,d_{4.9}^2\,{\rm erg\,s^{-1}}$. We have not found any clear evidence
of line emissions in the GIS spectra, and derive an equivalent width
upper limit of $EW<400$~eV for a narrow line at 6.7~keV from helium-like
iron. This equivalent width and the elemental abundance (table 1) are
not inconsistent with the characteristics found in white-dwarf
binaries (e.g., Ezuka, Ishida 1999). 
The remnant is in the crowded galactic plane, and it is not
surprising that a point-like source has been found within the same field of
view. We therefore conclude that G330.2+1.0 and AX~J1601$-$5143 are not physically
associated, and that G330.2+1.0 is a shell-type SNR without an energetic
pulsar.

From a statistical point of view alone, we can not distinguish the
better spectral model for the SNR shell (table
\ref{tab:spectral_parameter}). If the thermal model is appropriate,
the relatively high temperature ($kT=2.99$~keV) suggests that
the SNR is young and emission lines from metal-rich ejecta should
be expected.  The absence of significant Mg, Si, S, and Fe K shell
lines is incompatible with our knowledge for young, high-temperature
SNRs, such as Cas-A, Tycho's, or Kepler's (e.g., Tsunemi et al. 1986;
Kinugasa et al. 1999).  The suppression of emission lines could occur
if the interstellar medium density is very low, resulting in the low
ionization state. This is unlikely, since the clumpy radio morphology
suggests shock interaction with the dense medium.
Therefore, it is difficult to
explain the spectrum in the thermal model. We thus conclude that the
most reasonable interpretation for the featureless spectrum in G330.2+1.0 is
synchrotron radiation from relativistic electrons accelerated in
the SNR shell.

To evaluate multi-wavelength emission, we fit the SNR spectrum
with the {\tt srcut} model, which describes synchrotron radiation from
a power-law distribution of electrons with an exponential cut-off (Reynolds, Keohane 1999). For a fixed radio spectral index ($\alpha=0.3$)
and flux density (5 Jy at 1 GHz) (Green 2004), the fit is
significantly worse ($\chi^2/{\rm dof}=470.6/392$) than that of the simple power-law
model. This means that the radio and X-ray emission can not have
originated from a simple, single population of electrons in a uniform
magnetic field.  The best-fit parameters are the break frequency of
$\nu_{\rm b} = 4.3\times 10^{15}$~Hz and the absorption column of
$n_{\rm H}=5.1\times 10^{22}\, {\rm cm^{-2}}$. If the radio flux density is
considered as a free parameter, the fit becomes better
($\chi^2/dof=470.6/392$) with $\nu_b = 7.0\times 10^{16}$~Hz, a flux
density of $S=7.3\times 10^{-3}$~Jy, and $n_{\rm H}=2.2\times 10^{22}\, {\rm
cm^{-2}}$. The break frequency corresponds to an electron energy of
$E_{\rm b} = 30 \cdot (B_5)^{-1/2}$~TeV. Here, $B_5$, is the magnetic field
of the emitting region normalized to 5\, ${\rm \mu G}$.  The small
best-fit flux density ($7.3\times 10^{-3}$~Jy at 1\, GHz) compared to
that observed (5 Jy) suggests that there are at least two populations
of particles, and only a tiny fraction of the radio-emitting particles
has an extended tail toward the TeV region.

As can be seen in figure \ref{fig1}, the surface brightness of the X-ray
emission is anti-correlated with the radio emission.  The X-ray
emission is strongest at the western shell, where the lower radio
intensity toward the higher galactic latitude suggests a lower ISM
density. It is likely that moderately accelerated (GeV) electrons are
efficiently decelerated in the eastern shell interacting with the
dense ISM. In contrast, the lower ISM density in the western shell
results in efficient acceleration to the TeV range.  The
co-existence of two distinct (radio bright/X-ray faint and radio
faint/X-ray bright) shells in a single supernova remnant gives us an
interesting opportunity to study the acceleration mechanisms in
different ISM environments.

Interestingly, the X-ray luminosity of the shell, $L_{\rm SNR}=4.6\times
10^{34}\,d_{4.9}^2\,{\rm erg\,s^{-1}}$ (0.7--10~keV) is comparable to
that for RX~J1713.7$-$3946 ($1\times 10^{35}\,d_{1}^2$), SN~1006 ($\simeq 2\times
10^{35}\,d_{2.2}^2\, {\rm erg\,s^{-1}}$), RX~J0852.0$-$4622
($2\times10^{34}\,d_{1}^2\, {\rm erg\,s^{-1}}$) (Slane et al. 1999; Slane et
al. 2001), within a factor of $\simeq 5$.  There may be common physics
at work for tuning the acceleration efficiency in these SNRs.

In summary, we reported on the discovery of a featureless X-ray spectrum in
the SNR G330.2+1.0. We interpret this emission as coming from synchrotron
radiation of electrons accelerated at the SNR shock as seen in
SN~1006 and RX~J1713.7$-$3946. Our preliminary analysis of the archived XMM-Newton
data with short exposure ($\sim 9$~ks) qualitatively confirmed our
findings presented herein. Further analysis of the XMM-Newton data
with an emphasis on the nature of AX~J1601$-$5143 will be presented elsewhere.
Since the surface brightness of the shell is low, X-ray observations
with Suzaku, which has a large effective area and low background,
will be useful for an accurate determination of the SNR
spectrum. Although neither G330.2+1.0 nor the compact source was detected in
the recent H.E.S.S. survey (Aharonian et al. 2006), additional deep
searches for TeV emission may be useful. For constraining different
radiation mechanisms (inverse Compton effect, pion decay, non-thermal
bremsstrahlung) in the TeV region, radio observations of the
environmental molecular gas (e.g., Fukui et al. 2001) will be of
great help.

\section*{}
The authors are grateful to the referee, P. Slane, for valuable
suggestions that greatly improved the original draft.  

This work is
partly supported by a Grant-in-Aid for Scientific Research by the
Ministry of Education, Culture, Sports, Science and Technology (16002004). This work was financially supported by the Japanese
Ministry of Education and the 21st Century COE Program named "Towards
a New Basic Science: Depth and Synthesis".

\end{document}